\documentclass[prd,superscriptaddress,showpacs,twocolumn,nofootinbib,floatfix,amsmath,amssymb]{revtex4-1}
\usepackage{mathtools}
\usepackage{mathrsfs}
\usepackage[usenames,dvipsnames]{xcolor}
\usepackage{color}
\usepackage{braket}
\usepackage{bm}
\usepackage{graphicx}
\usepackage{hyperref}
\usepackage{epstopdf}

\begin{document}

\title{Ideal clocks - a convenient fiction}

\author{Krzysztof Lorek}
\affiliation{Institute of Theoretical Physics, Faculty of Physics, University of Warsaw, Pasteura 5, 02-093 Warsaw, Poland}

\author{Jorma Louko}
\affiliation{School of Mathematical Sciences, University of Nottingham, Nottingham NG7 2RD, United Kingdom}

\author{Andrzej Dragan}
\affiliation{Institute of Theoretical Physics, Faculty of Physics, University of Warsaw, Pasteura 5, 02-093 Warsaw, Poland}

\begin{abstract}
We show that no device built according to the rules of quantum field theory can measure proper time along its path. Highly accelerated quantum clocks experience the Unruh effect, which inevitably influences their time rate. This contradicts the concept of an ideal clock, whose rate should only depend on the instantaneous velocity.
\end{abstract}

\maketitle

One immediate prediction of special relativity is that a rate of any clock moving inertially with a velocity $v$ is dilated by a Lorentz factor $\sqrt{1-v^2}$ (we work in natural units such that: $\hbar=c=1$) independent of the clock's mechanism. The same law applies to a pendulum clock and an atomic clock, as according to the principle of relativity motion of any of these clocks is equivalent to the motion of the observer in the opposite direction. Special relativity cannot predict however how an arbitrary clock is affected by non-inertial motion, because different clocks, for example the pendulum clock or the atomic clock, will be affected by acceleration in a different way. One often introduces a {\em clock postulate} defining a hypothetical {\em ideal clock} as a device that measures proper time $\tau$ along its arbitrary path according the formula \cite{Wald2003}:
\begin{equation}
\label{idealclock}
\tau = \int_{\text{path}} \sqrt{1-v^2(t)}\,\text{d}t,
\end{equation}
which does not depend on the clock's acceleration at all, only on its instantaneous velocity $v(t)$. The assumption \eqref{idealclock} in its idealized form leads to interesting consequences. Consider an ideal clock oscillating along a sinusoidal path: $x(t) = A\sin\omega t$, where $A$ and $\omega$ are given amplitude and frequency, respectively. The clock's velocity and acceleration vary according to: $v(t) = A\omega\cos\omega t$, $a(t) = -A\omega^2\sin \omega t$. Let us consider a limit of small amplitudes and high frequencies such that $A\to 0$, $A\omega\to 0$ and $A\omega^2 \to \infty$. In this non-relativistic limit the clock remains at rest at $x=0$ with no velocity $v=0$ but with divergently oscillating acceleration. According to the ideal clock formula \eqref{idealclock} the rate of such a clock will be exactly the same as the rate of a normally resting clock. This conclusion may go against one's intuition that singular accelerations should somehow affect the clock rate, since all physical devices, such as the simplest pendulum clock are clearly affected by acceleration.

It is often argued that proper time \eqref{idealclock} offers the perfect description of an ideal clock, because it remains invariant under Lorentz transformations. This is an obvious requirement for clocks or any other physical measurement devices, because one expects them to measure quantities that do not depend on the choice of the observer. Proper time along the path however is not the only possible invariant that characterizes the classical trajectory. Consider a generic path $x(t)$ and take the four-acceleration $a^\mu$ characterized by an instantaneous velocity $v(t) = \frac{\text{d}x}{\text{d}t}$ and its derivative $a(t) = \frac{\text{d}v}{\text{d}t}$: 
\begin{equation}
a^\mu = \left( \frac{av}{(1-v^2)^2}, \frac{a}{(1-v^2)^2}  \right).
\end{equation}
One can construct the following example of an alternative invariant characterizing the path by integrating the length of the above four-vector over proper time:
\begin{eqnarray}
\text{invariant} &=& \int_{\text{path}} \sqrt{-\eta_{\mu\nu}a^\mu a^\nu} \text{d}\tau = \int_{\text{path}} \alpha(\tau)\text{d}\tau \nonumber \\ \nonumber \\
 &=& \int_{\text{path}}  \frac{a(t)}{1-v^2(t)}\text{d}t,
\end{eqnarray}
where $\eta_{\mu\nu}$ is Minkowski metric and $\alpha$ is proper acceleration. The above dimensionless invariant, as well as its higher-order alternatives could be easily used to model hypothetical acceleration dependent corrections to the ideal clock formula \eqref{idealclock}.

In this work we investigate a general question, whether ideal clocks may exist in nature, i.e. whether known laws of physics allow one, even in principle, to build a device that measures proper time \eqref{idealclock} along every path. In a recent work \cite{Lindkvist2014} it has been shown that a finite-size clock based on the interference effect of motion along two different paths shows deviations from the ideal clock formula \eqref{idealclock}. Experimentally, no deviation from the ideal clock formula has been found so far for clocks based on a decay time of unstable particles \cite{BaileyN,Bailey,Botermann,Mainwaring}. Here we show that rates of such clocks must inevitably deviate from the ideal clock formula \eqref{idealclock}. Moreover, these deviations cannot be compensated for if the device is to measure the proper time along an arbitrary space-time trajectory. We shall work within quantum field theory on a fixed background spacetime \cite{Birrell1984} and leave connections to generally covariant theories \cite{Rovelli} subject to future work.

First we need to ask what is the most fundamental clock within the framework of quantum field theory in non-inertial frames or curved spacetimes. Since all the time-scales of dynamical processes in quantum field theory ultimately boil down to the fundamental time-scales of particle interactions, it is reasonable to start with an observation that the most fundamental clock one can think of would be based on a decay of an unstable particle. Such a clock would measure time in terms of the lifetimes of a given standard particle \cite{BaileyN,Bailey}. Ultimately all time-scales known in Nature are derived from these decay rates. The existence of the effect of accelaration on such decaying particles, is supported by models that have been studied previously \cite{Mueller}.

Let us start with a few observations. Typical states of particles encountered in reality are described with localized wave-packets. If we intend to use fundamental particles as model clocks they are bound to have a finite size. Trying to localize particles in an increasingly smaller spatial region eventually leads to particle-antiparticle pair creation, which could affect the decay rate. If a wave-packet containing an unstable particle is moving with an acceleration, one can note that different parts of the wave-packet move along slightly different trajectories, and are characterized by slightly different proper accelerations and proper times. What would be the overall effect of this spread on the lifetime of the particle? Another observation is that the accelerating object experiences the surrounding Minkowski vacuum as a thermal state of a temperature proportional to the proper acceleration due to Unruh effect \cite{Unruh1976}. It is easy to imagine that an interaction with such a thermal state will inevitably affect the particle decay rate, and our results indeed confirm this.

In our work we model the clock as a decaying particle localized within the lowest energy mode of a finite-sized, one-dimensional cavity confining a massless scalar quantum field. Using a finite-sized cavity instead of introducing a wave-packet description allows us to neglect the effects of particle spreading and simplifies calculations without compromising the results. In order to facilitate the decay we introduce an external massive\footnote{The goal of the work was to demonstrate the effect of the simplest possible case. However, a massless external field was found to lead to infrared divergences, which could have been dealt with using renormalization, yet the authors believe that these would only obscure the simplicity and the generality of our result, which is independent of technical details such as e.g. choosing a renormalization scheme.} scalar quantum field initially in the vacuum state and consider the simplest model of interaction between the two fields. As a result of this interaction, the considered cavity state has a tendency to decay into its overall ground state accompanied by an excitation an of the external field. This decaying cavity particle is the most fundamental model of a quantum clock with the decay rate corresponding to the clock's ticking rate. In order to simulate the motion of the clock, we calculate the decay rates for two scenarios: a stationary clock corresponding to the cavity at rest and a uniformly accelerated clock corresponding to the cavity moving with a uniform relativistic acceleration. The second scenario also corresponds to an equivalent case of a clock placed in a static gravitational field. Our calculation shows that the decay rates are affected by the acceleration, however the special-relativistic time dilation is not the only effect responsible for the change. We show that other processes also inevitably occur affecting the outcomes in a way that is in disagreement with the formula \eqref{idealclock}. Interestingly, we do not retrieve the exact special relativistic formula even in the limiting case when the length of the clock cavity goes to zero. Such correspondence is obtained only for low accelerations.

The discrepancy with the prediction of the formula \eqref{idealclock} can be understood from the point of view of the accelerated clock's reference frame. In this frame the clock (or cavity) is stationary but no longer interacts with the vacuum of the external field. Instead, as mentioned before, it perceives the state of the external field as being in a thermal state. The presence of surrounding particles modifies the decay law, which has been shown in this work.

In our study we consider a $1+1$ dimensional cavity of a proper length $l$ with Dirichlet boundary conditions. The cavity can either stay at rest or uniformly accelerate. The inertial lab-frame coordinates of the cavity walls at $t=0$ for both cases are: $\sigma_-,\sigma_+$. Throughout the paper we keep the consistent notation that quantities characterizing the external field are capitalized, and those referring to the cavity field are not.

The two Klein-Gordon scalar fields under consideration are the cavity massless field $\hat{\phi}$ and a massive external field $\hat{\Phi}$ of a strictly positive mass $M$, occupying the volume both inside and outside the cavity. We assume that the cavity walls are transparent to the external field and assume that the system of the two fields is subject to the simplest possible coupling\footnote{Using other types of fields and couplings does not change qualitative conclusions presented here. One can also study a semi-classical model of coupling via Unruh-DeWitt Hamiltonian and similar conclusions can be drawn using such approach.}, described by following interaction Hamiltonian:
\begin{equation} \label{HInt}
\hat{H}_{int}=\lambda\int\text{d}x\,\hat{\phi}\,\hat{\Phi},
\end{equation}
where $\lambda$ is a small coupling strength.

Let us introduce the following decomposition of the scalar fields:
\begin{align}
\hat{\phi}(x,t)={\sum\limits_n} \hat{a}_n u_n(x,t)&+\hat{a}_n^\dagger u_n^*(x,t),\nonumber \\
\hat{\Phi}(x,t)={\int_K} \hat{A}_K U_K(x,t)&+\hat{A}_K^\dagger U_K^*(x,t),
\end{align}
where $\hat{a}$'s and $\hat{A}$'s denote annihilation operators of the cavity field and the external field, respectively and their corresponding field modes are $u$'s and $U$'s. We can now proceed with the calculation of the decay rate of a one-particle cavity excitation of the mode $u_1$ and the ground state of the external field into the cavity ground state and an arbitrary final state of the external field $|\beta\rangle$. We first consider the simplest scenario when the massless field cavity is at rest. For simplicity we limit ourselves to the first order perturbation theory, in which the decay amplitude $\mathscr{A}_\downarrow$ is given by: 
\begin{equation}
\mathscr{A}_\downarrow=-i\,\int_0^t\text{d}t' \bra{0}_\phi\bra{\beta}_\Phi \hat{H}_{int}\ket{1}_\phi\ket{0}_\Phi
\end{equation}
and after summing over all possible final states of the external field we obtain the corresponding probability of the decay $P_\downarrow$:
\begin{equation} \label{PDownS}
P_\downarrow =\lambda^2\int_K\left|\gamma_{K1}\right|^2,
\end{equation}
where $\gamma_{Kn}$ denotes the following time-integrated mode overlap:
\begin{equation}
\gamma_{Kn}=\int_0^t\text{d}t'\int\text{d}x\,U_K^*u_n.
\end{equation}
For the considered case of the stationary cavity clock the field modes are simply:
\begin{equation}
u_n(x,t)=\frac{1}{\sqrt{\pi n}}\sin\big(\omega_n(x-\sigma_-)\big)e^{-i\omega_n t}
\end{equation}
inside the cavity and zero outside, where: $\omega_n=\frac{n\pi}{l}$, $l=\sigma_+-\sigma_-$. The external field is decomposed into plane waves:
\begin{equation}
U_K(x,t)=\frac{1}{\sqrt{4\pi\Omega_K}}e^{iKx-i\Omega_K t},
\end{equation}
where $\Omega_K=\sqrt{K^2+M^2}$. The above formulas can be substituted into \eqref{PDownS} giving the following expression:
\begin{align} \label{PSCalka}
P_\downarrow=
\frac{4\lambda^2}{l^2}{\int\limits_{-\infty}^{+\infty}}\text{d}K
\frac{\cos^2\left(\frac{Kl}{2}\right)\sin^2\left((\Omega_K-\frac{\pi}{l})\frac{t}{2}\right)}
{(\Omega_K-\frac{\pi}{l})^2(K^2-\frac{\pi^2}{l^2})^2\,\Omega_K}.
\end{align}
For sufficiently short times the sin function can be replaced with its argument and the expression becomes proportional to $t^2$. In the long time limit $t\rightarrow\infty$ the expression $\sin^2 (xt)/x^2t$ becomes proportional to Dirac's delta $\delta(x)$ and the integration can be approximated by:
\begin{equation} \label{PSM}
P_\downarrow=
\frac{4\lambda^2\pi t\cos^2\left(\sqrt{\frac{\pi^2}{l^2}-M^2}\frac{l}{2}\right)}
{l^2M^4\sqrt{\frac{\pi^2}{l^2}-M^2}},
\end{equation}
provided that $\frac{\pi}{l}>M$, and zero otherwise. This has the following physical interpretation: in the limit of $t\rightarrow\infty$ energy in the decay is conserved therefore the transition must be resonant. If we now let either $l\rightarrow 0$ or $M\rightarrow 0$, in both cases the leading term is:
\begin{align} \label{PSMzero}
P_\downarrow=
\frac{\lambda^2l^3t}{4\pi^2},
\end{align}
thus we see that when our clock is small, the probability of the decay does not depend on the mass of the free field.

Let us now proceed with a more complex problem, where the cavity mirrors follow trajectories of a  relativistic, uniformly accelerated motion simulating the behavior of a uniformly accelerated clock. This analysis becomes substantially simplified, when analyzed from the uniformly accelerated reference frame co-moving with the cavity. Such a system is conveniently described using Rindler coordinates $(\tau,\xi)$:
\begin{equation}
\label{RindlerTrans}
\begin{split}
\xi=&\frac{1}{\alpha}\ln\left(\alpha\sqrt{x^2-t^2}\right)\\
\tau=&\frac{1}{\alpha}\text{arctanh}\left(\frac{t}{x}\right),
\end{split}
\end{equation}
where $\alpha$ is the proper acceleration of the reference trajectory and the temporal coordinate $\tau$ is chosen such to be the proper time along that reference trajectory. We choose the reference trajectory to be exactly in the center of the cavity, such that $\alpha= \frac{2}{\sigma_-+\sigma_+}$ refers to the average proper acceleration of the cavity, thus $\tau$ is approximately the proper time of the considered localized particle. On the technical side, let us notice that the above coordinates cover only the right Rindler space-time wedge $\text{I}$ given by $x>|t|$. By mirror reflecting the spatial coordinate in the Rindler transformation one can also cover the left Rindler wedge $\text{II}$ with an analogous set of coordinates and eventually decompose the field operators at $t = \tau = 0$ into a complete set of Rindler modes \cite{Birrell1984}:
\begin{align}
\nonumber
\hat{\Phi}(\xi,\tau)=\int_K &\hat{B}_{K,\text{I}} U_{K,\text{I}}(\xi,\tau)+\hat{B}_{K,\text{I}}^\dagger U_{K,\text{I}}^*(\xi,\tau)+\\+
&\hat{B}_{K,\text{II}} U_{K,\text{II}}(\xi,\tau)+\hat{B}_{K,\text{II}}^\dagger U_{K,\text{II}}^*(\xi,\tau) \\ \nonumber \\
\hat{\phi}(\xi,\tau)={\sum\limits_n} &\hat{b}_n u_n(\xi,\tau)+\hat{b}_n^\dagger u_n^*(\xi,\tau),\nonumber
\end{align}
where the $\text{I/II}$ index of the $\hat{B}_{K,*}$ and $U_{K,*}$ Rindler modes stands for the right and left Rindler wedge and the modes $U_{K,*}$ are solutions to the Klein-Gordon equation in Rindler coordinates. The massless field equation is conformally invariant under Rindler transformation \eqref{RindlerTrans}, hence the mode solutions $u_n(\xi,\tau)$ in the accelerated frame have exactly the same form as in the rest frame $u_n(x,t)$.

The accelerated cavity moves through the Minkowski vacuum $|0\rangle_{\text{M}}$ state of the external field, which in the Rindler reference system has a complicated structure that involves squeezing $\hat{S}_{\text{I,II}}$ of the two Rindler wedges:
\begin{equation}
|0\rangle_{\text{M}} = \hat{S}_{\text{I,II}} |0\rangle_{\text{R}},
\end{equation}
where $|0\rangle_{\text{R}}$ is the Rindler vacuum. In particular this state reduced to only one of the wedges is exactly a thermal state, which is known as the Unruh effect. We choose the initial state of the cavity field to be a single Rindler particle of the lowest energy cavity mode, so again the cavity contains a single particle from the perspective of a co-moving observer. Following the previous analysis we calculate the first-order perturbation expansion of the evolution operator and calculate the decay probability amplitude:
\begin{equation}\label{decayamplrind}
\mathscr{A}_\downarrow=-i\,\int_0^\tau\text{d}\tau'\bra{0}_\phi\bra{\beta}_{\Phi} \hat{S}^\dagger_{\text{I,II}}\hat{H}_{int}\ket{1}_\phi \hat{S}_{\text{I,II}}\ket{0}_{\Phi},
\end{equation}
where all the bras and kets are the states defined with respect to the Rindler reference frame. The description is made in the accelerated frame, therefore we will use the Hamiltonian $\hat{H}_{int}$ in a constant $\tau$ foliation.

In order to explicitly calculate the decay amplitude \eqref{decayamplrind} one needs to commute the interaction Hamiltonian $\hat{H}_{int}$ and the squeezing operator $\hat{S}_{\text{I,II}}$. This can be easily done using the commutation relations characterizing the squeezing operator $\hat{S}_{\text{I,II}}$ \cite{Dragan2013,Takagi}:
\begin{equation}
\hat{S}_{\text{I,II}}^\dagger \hat{B}_{K,\text{I}}\hat{S}_{\text{I,II}}=\cosh r_{\Omega_K} \hat{B}_{K,\text{I}}+\sinh r_{\Omega_K} \hat{B}_{K,\text{II}}^\dagger,
\end{equation}
where $r_{\Omega_K}$ is the acceleration-dependent squeezing parameter satisfying the equation: $\tanh r_{\Omega_K} = e^{-\frac{\pi\Omega_K}{\alpha}}$.

The calculation then proceeds in analogous fashion as in the stationary clock case to yield the following formula for the probability of the decay:
\begin{equation}
P_\downarrow = \lambda^2{\int_K}
\left(\left|\gamma_{K1}\right|^2 + \sinh^2r_{\Omega_K}
\left(\left|\gamma_{K1}\right|^2 + \left|\bar{\gamma}_{K1}\right|^2\right)
\right),
\end{equation}
where now the following overlaps arise:
\begin{eqnarray}
\nonumber
&\gamma_{Kn}=\int_0^\tau\text{d}\tau'\int\text{d}\xi\,U_{K,\text{I}}^*u_n
\\
&\bar{\gamma}_{Kn}=\int_0^\tau\text{d}\tau'\int\text{d}\xi\,U_{K,\text{I}}u_n.
\end{eqnarray}
The differences from the formula (\ref{PDownS}), which describes the stationary case, stem from the non-trivial transformation properties of quantum states due to acceleration. We can explicitly evaluate the above formulas by inserting to the above formulas the cavity modes:
\begin{equation}
u_n(\xi,\tau)=\frac{1}{\sqrt{\pi n}}\sin\big(\omega_n(\xi-\xi_-)\big)e^{-i\omega_n \tau},
\end{equation}
where: $\omega_n=\frac{\alpha n\pi}{\ln\frac{\sigma_+}{\sigma_-}}$, and $\xi_\pm=\frac{1}{\alpha}\ln(\alpha\sigma_\pm)$ are the Rindler positions of the cavity walls. To complete the calculation we also need the external field modes in the Rindler frame, which can be calculated as follows. The free field is now governed by the Klein-Gordon equation transformed to the Rindler frame:
\begin{equation} \label{KGAcc}
\big(\,\frac{\partial^2}{\partial\tau^2}-\frac{\partial^2}{\partial\xi^2}+
M^2e^{2\alpha\xi}\,\,\big)\Phi(\xi,\tau)=0,
\end{equation}
with the solutions of the form:
\begin{equation}
U_\Omega(\xi,\tau)=F_\Omega(\xi)e^{-i\Omega\tau},
\end{equation}
where the spatial functions $F_\Omega(\xi) $ satisfy the modified Bessel equation. Hence the explicit formula contains a modified Bessel function of the second kind, ${\cal K}$ \cite{Takagi}:
\begin{equation} \label{AccMode}
F_\Omega(\xi)=\frac{1}{\sqrt{\pi\Omega}}
\Big(\frac{M}{2\alpha}\Big)^{\frac{i\Omega}{2\alpha}}
\frac{1}{\Gamma(\frac{i\Omega}{\alpha})}
{\cal K}_{\frac{i\Omega}{\alpha}}\left(\frac{M}{\alpha}e^{\alpha\xi}\right),
\end{equation}
From here we can write down explicitly the probability of the decay of the accelerating particle:
\begin{widetext}
\begin{eqnarray}
\label{pdownacc}
P_\downarrow=&\nonumber
\frac{4\lambda^2}{\pi^2}
{\int\limits_{0}^{+\infty}}\frac{\text{d}\Omega}{\Omega}
\frac{1}{\left|\Gamma(\frac{i\Omega}{\alpha})\right|^2}
\left|\,
{\int\limits_{\xi_-}^{\xi_+}}\text{d}\xi\,
{\cal K}_{\frac{i\Omega}{\alpha}}\left(\frac{M}{\alpha}e^{\alpha\xi}\right)
\sin\left(\omega_1\left(\xi-\xi_-\right)\right)
\right|^2 \Bigg[
\frac{\sin^2\left[\left(\Omega-\omega_1\right)\frac{\tau}{2}\right]}
{\left(\Omega-\omega_1\right)^2}
+
\frac{1}{e^{\frac{2\pi\Omega}{\alpha}}-1}
\Bigg(
\frac{\sin^2\left[\left(\Omega-\omega_1\right)\frac{\tau}{2}\right]}
{\left(\Omega-\omega_1\right)^2}
+
\frac{\sin^2\left[\left(\Omega+\omega_1\right)\frac{\tau}{2}\right]}
{\left(\Omega+\omega_1\right)^2}
\Bigg)\Bigg].
\end{eqnarray}
\end{widetext}
which is the final expression for the decay probability of the accelerated cavity that can be directly compared with the stationary case \eqref{PSCalka}. For long interaction times it simplifies to
\begin{equation} \label{PAccM}
P_\downarrow=\frac{\lambda^2\tau e^{\frac{\pi\omega_1}{\alpha}}}{\pi^2\alpha}
\left|\,\,
{\int\limits_{\xi_-}^{\xi_+}}\text{d}\xi\,
{\cal K}_{\frac{i\omega_1}{\alpha}}\left(\frac{M}{\alpha}e^{\alpha\xi}\right)
\sin\left(\omega_1\left(\xi-\xi_-\right)\right)
\right|^2.
\end{equation}

Let us now see in which limiting cases the formula derived above corresponds to the stationary clock case. It can be shown that the above result has the form of a smooth envelope with superimposed oscillations of the frequency diverging at $\alpha\rightarrow 0$. The oscillations are a subtle consequence of the boundary conditions imposed in Rindler coordinates. For weakly accelerated cavity one can get rid of these oscillations by averaging the result around the chosen value $\alpha$, which corresponds to a finite uncertainty about the value of acceleration. We perform this averaging for $\frac{\pi}{l}>M$ \cite{Dunster, Functions} in the limit of small $\alpha$, keeping the cavity size $l$ fixed, to obtain:
\begin{equation} \label{PAM}
P_\downarrow=
\frac{4\lambda^2\pi\tau\cos^2\left(\sqrt{\frac{\pi^2}{l^2}-M^2}\frac{l}{2}\right)}
{l^2M^4\sqrt{\frac{\pi^2}{l^2}-M^2}}.
\end{equation}
In this limit we retrieve the formula \eqref{PSM} with $t$ replaced with $\tau$, which corresponds to the ideal clock case. For large accelerations $\alpha$, however, a strong discrepancy between the accelerated cavity and resting cavity cases arises. It is clear that the expressions \eqref{PSM} and \eqref{pdownacc} are different and cannot be related via simple substitution $t\to\tau$. These differences can be related to the Unruh effect, whose importance rises as the proper acceleration of the cavity increases. One might only expect that the ideal clock formula can work for decaying particles with accelerations sufficiently small. Obviously, if the clock's trajectory is known in advance, one can artificially compensate for the clock rate difference. The purpose of an ideal clock is however, to measure time on every trajectory without any prior knowledge of its shape. In this case no compensation is possible and any device based on the physical principles captured by our model will measure the proper time only approximately and for small proper accelerations compared to the typical proper accelerations at which the Unruh effect can be observed. This regime is still many orders of magnitude above the everyday accelerations, that's why it is perfectly possible to build time-measuring device working well for typical everyday accelerations experienced on Earth. However, the accelerations at which one expects Unruh effect to be detectable are becoming empirically accessible, hence further work will be undertaken to investigate a more realistic setup, and provide a definite experimental prediction for the effect.

The result that the time measured by a clock, depends on its acceleration, has been derived here for the simplest case of a uniform acceleration. However, if this effect exists in this case, it will naturally persist for a general accelerated motion, which the one considered here is a special case of. Returning to the divergently oscillating clock paradox, described in the introduction, the authors suspect that indeed its reading would deviate from the prediction of eq. \eqref{idealclock}. Nevertheless considering different types of accelerated motion may lead to interesting new results. Similarly one could also think of investigating the effect on more complicated fields, e.g. a spinor field. Qualitatively the behaviour is expected to be the same, but it is definitely worth further studying.

Ideal clock formula describing a device fundamentally insensitive to the experienced proper accelerations is a convenient, but fictitious concept. All known physical processes and consequently all devices must become sensitive to their accelerations at certain scales and therefore the rate of any physical clock must inevitably differ from the idealized formula \eqref{idealclock}. 

\section*{Acknowledgements}

Authors would like to thank Ivette Fuentes, Krzysztof Pachucki, and Jan Derezi\'{n}ski for useful comments and discussions. J.~L. was supported in part by STFC (Theory Consolidated Grant ST/J000388/1). A.~D. thanks for the financial support to the National Science Center, Sonata BIS Grant No. 2012/07/E/ST2/01402.

\end{document}